\RequirePackage{lineno}
\documentclass[superscriptaddress,amsmath,amssymb,nofootinbib,twocolumn]{revtex4-2}

\usepackage{graphicx}
\usepackage{dcolumn}
\usepackage{bm}
\usepackage{ulem}
\usepackage{xfrac}
\usepackage{lipsum}
\usepackage{braket}
\usepackage{times}
\usepackage[dvipsnames]{xcolor}

\usepackage{graphicx,amsmath,amssymb}
\usepackage{epstopdf}
\usepackage{lineno}
\usepackage[colorlinks=true, urlcolor=blue, linkcolor=blue, citecolor=blue]{hyperref}

\begin{document}

\title{$\,$\\[-6ex]\hspace*{\fill}{\normalsize{\sf\emph{Preprint nos}.\
JLAB-PHY-26-4853;\;
SLAC-PUB-260704;\; NJU-INP 120/26}}\\[1ex]
Regarding the Rotational Unruh Effect} 

\author{Alexandre~Deur%
$\,^{\href{https://orcid.org/0000-0002-2203-7723}{\textcolor[rgb]{0.00,1.00,0.00}{\sf ID}}}$}
\affiliation{Thomas Jefferson National Accelerator Facility, Newport News, Virginia 23606, USA }

\author{Stanley J. Brodsky%
$\,^{\href{https://orcid.org/0000-0001-8786-3172}{\textcolor[rgb]{0.00,1.00,0.00}{\sf ID}}}$}
\affiliation{SLAC National Accelerator Laboratory, Stanford University, Stanford, California 94309, USA}

\author{Craig D.~Roberts%
$\,^{\href{https://orcid.org/0000-0002-2937-1361}{\textcolor[rgb]{0.00,1.00,0.00}{\sf ID}}}$}
\affiliation{School of Physics, Nanjing University, Nanjing, Jiangsu 210093, China}
\affiliation{ Institute for Nonperturbative Physics, Nanjing University, Nanjing, Jiangsu 210093, China}

\author{Bal{\v s}a Terzi{\'c}%
$\,^{\href{https://orcid.org/0000-0002-9646-8155}{\textcolor[rgb]{0.00,1.00,0.00}{\sf ID}}}\,$}
\affiliation{Department of Physics, Old Dominion University, Norfolk, Virginia 23529, USA}

\date{2026 July 12}

\begin{abstract}
\centerline{Email:
\href{mailto:deurpam@jlab.org}{deurpam@jlab.org} (AD);
\href{mailto:sjbth@slac.stanford.edu}{sjbth@slac.stanford.edu} (SJB);
\href{mailto:cdroberts@nju.edu.cn}{cdroberts@nju.edu.cn} (CDR);
\href{mailto:bterzic@odu.edu}{bterzic@odu.edu} (BT)}
\,\\[1ex]
 We study the rotational Unruh effect by identifying the pseudodynamics which emerges when fundamental spacetime symmetries are violated. We show that there is, fundamentally, no rotational Unruh effect, in agreement with the first explicit calculations of the literature, and that, when such an effect has been claimed to exist in later articles, its origin traces back to the derivation of the standard Unruh effect involving linear acceleration. Such a linear Unruh effect intrudes because the chosen reference frames do not respect the cylindrical symmetry of the system, and thus the analyses require Lorentz boosts. Our study also confirms previous findings that the Unruh effect is a consequence of the pseudodynamics which arises because Instant-Form dynamics explicitly breaks a fundamental symmetry of spacetime: Poincar\'e invariance.
 
\end{abstract}

\maketitle

\section{Introduction}

The Unruh effect is the prediction that an observer under uniform acceleration will observe the materialization of the virtual particles which are often supposed to populate the ``vacuum'' of a quantum field theory (QFT).
This effect would be observed via the heating of the accelerating detector to a temperature proportional to its acceleration \cite{Fulling:1972md, Davies:1974th, Unruh:1976db, Crispino:2007eb}. 
An issue with this prediction is that if a QFT is analyzed in a manner that explicitly preserves Poincar\'e invariance, then the vacuum is bereft of virtual particles. Consequently, in such an analysis, an Unruh effect is impossible \cite{Deur:2024szw}.

Poincar\'e invariance is the basic symmetry of spacetime and a foundation for QFT. 
Fields satisfying Poincar\'e invariance may be quantized in different ways, depending on the convention adopted for foliating spacetime into space and time \cite{Dirac:1949cp}.   
Dirac's Front-Form (FF) quantization of a field \cite{Brodsky:1997de}, where the time $x^+$ and one space direction $x^{-}$ are defined as being along a light cone, explicitly preserves Poincar\'e invariance for all intents and purposes \cite{Brodsky:2022fqy}. 
A straightforward consequence of FF quantization is the triviality of the vacuum, \textit{viz}.\ it is entirely empty, with zero virtual particles.

On the other hand, the other common/conventional quantum field theory  quantization scheme, Instant-Form (IF), where time is defined as the usual Galilean time $t$, violates Poincar\'e invariance. 
To compensate for this violation, pseudodynamics become necessary, including the introduction of a nontrivial, virtual-particle-full vacuum. 

Physically, IF and FF analyses must lead to the same conclusions, since truly observable physical cannot not depend on convention choices, like the manner in which spacetime is foliated or what reference frame is chosen for the analysis. 
However, this entails that any IF analysis of a given phenomenon must include all pseudoeffects necessary to remedy the inherent symmetry violation.  

The following analogy may be useful.
The basic space symmetry in classical Newtonian dynamics is Galilean invariance. 
One is free to choose to analyze a problem in a frame preserving Galilean
invariance, \textit{viz}.\ an inertial frame, or in a non-inertial frame, where Galilean invariance is violated. 
The loss of this symmetry, which entails the loss of momentum conservation, is accommodated by the appearance of the centrifugal, Coriolis, and Euler pseudoforces. 
Conclusions from analyses performed in inertial and non-inertial frames will agree, unless the pseudoforces are overlooked.  

The reason for the apparently contradictory conclusions about the Unruh effect in FF and IF analyses is thus clear: the Unruh effect 
appears in IF dynamics because it does not respect Poincar\'e invariance. However, since not all pseudodynamics has been included, the effect appears to be objective and observable \cite{Deur:2024szw}. 
Using again the classical dynamics analogy: an incomplete analysis in a non-Galilean frame, in which some pseudoforces are overlooked, would yield an erroneous conclusion and force the imprudent practitioner into asserting that there must exist a new real force. 

The Unruh effect was originally derived within IF dynamics with linear acceleration.  
However, a body in a circular orbit is also subjected to an acceleration, although centripetal, with its tangential (linear) acceleration null and its speed constant. 
It is currently accepted in the literature, which relies on IF dynamics--except for Ref.\cite{Deur:2024szw}--that there is also a rotational Unruh effect (RUE); but reaching this conclusion has not been straightforward, with ambiguous conclusions depending on the particulars of the IF analysis \cite{Fulling:2014wzx}. 

However, it is important to know if the RUE occurs, not only for the theory but also at a practical level. 
Indeed, reaching linear accelerations of a magnitude sufficient for a feasible Unruh test is beyond our present capabilities, and it is not known if it would ever be possible to build a detector that could sustain 
such an acceleration without drowning any Unruh signal in much larger peripheral effects. 
In contrast, centripetal accelerations that would yield high Unruh 
temperatures have long been achieved in particle colliders, 
thus enabling the possibility of a practical Unruh test in this case. 

Herein, we analyze the RUE within both IF and FF dynamics, profiting from insights drawn elsewhere \cite{Deur:2024szw}; namely, that in the IF case the effect is pseudodynamical, arising from the violation of a fundamental spacetime symmetry.

\section{Spacetime symmetries and the Unruh effect \label{spacetime-sym-and-Unruh-effect}}
Poincar\'e invariance is associated with the 10 symmetries of spacetime, corresponding to 10 generators:
4 linear momenta for spacetime translations,
3 angular momenta for space rotations,
and 3 Lorentz boosts. 
To each generator is associated an operator, {\it e.g.},  the time translation operator is the Hamiltonian. 
Operators that do not involve ``time'' are called {\it kinematical operators} and constitute the stability group of the dynamical form. 
The other operators are called {\it dynamical operators}.   
In the IF case, the stability group has 6 generators: the 3 linear and 3 angular momenta. The 3 boosts and the Hamiltonian are the 4 dynamical operators.
The FF stability group is larger, with 7 generators: the linear momenta, boosts, and one angular momentum. 
The 3 dynamical operators are the other two angular momenta and the Hamiltonian. 
Crucially, in particle physics, systems typically display only one rotational symmetry, the one around the particle propagation direction. Thus, the two dynamical angular momenta are irrelevant to the problem and, for all intents and purposes, FF dynamics preserves Poincar\'e invariance. In contrast, boosts are always relevant in particle physics, thereby complicating IF analyses with pseudodynamics, including the Unruh pseudoeffect. 

The pseudodynamics responsible for the Unruh effect originates from the dynamical nature of the IF Lorentz boost operators, which entails a violation of Poincar\'e invariance. 
In contrast, FF dynamics avoids this problem because 
FF boosts are kinematical operators; 
thus the Unruh effect does not arise in FF dynamics, consistent with its simple vacuum.\footnote{The FF vacuum does contain zero-modes, \textit{i.e}., virtual loops of  zero 4-momentum particles; but possessing zero momentum, they cannot heat an Unruh detector, so can be ignored here \cite{Deur:2024szw}.}
A corollary is that IF operators that respect Poincar\'e invariance should also not produce pseudodynamics. 
A correct/physically-consistent IF analysis must include all pseudoeffects that arise from dynamical boosts and affect the system, including the Unruh detector. 

The spacetime symmetry relevant to the study of the RUE is rotational invariance around the axis of rotation. 
Crucially, IF angular momenta are kinematical operators, and thus do not produce pseudodynamics. Answering the question of the existence of a RUE is therefore trivial from symmetry considerations: there should be no RUE in an IF analysis. 

The same conclusion is also immediate in the FF analysis, thus consistent with IF dynamics, as it must be. 
Even though only one angular momentum operator is kinematical in FF dynamics, this is sufficient because, for a circularly orbiting particle, the system displays cylindrical symmetry, typically chosen to be the $z$-axis. 
This axis can also be chosen to define the FF coordinates: $x^{\pm} \equiv t \pm z$. 
Then, in the FF, rotation around $z$ is a kinematical operation; so, there is also no RUE.

Indeed, calculations relating to the RUE -- all performed with IF -- have long shown \cite{Denardo:1978dj, Letaw:1979wy} that the inertial (Minkowski) and accelerated (Rindler) frames have identical vacua, which signifies at face value that there is no RUE. 
This is consistent with the reasoning based on Poincar\'e invariance discussed above and in Ref.\,\cite{Deur:2024szw}. 
It also demonstrates the power of symmetry considerations by providing an immediate answer, so straightforward it is reached without needing calculations.

However, some practitioners nevertheless maintain that there actually is a RUE. 
This perception originates from the study of Bell and Leinaas \cite{Bell:1982qr, Bell:1986ir}. 
Ref.\,\cite{Letaw:1979wy} assessed that although the Minkowski and Rindler vacua are identical, an Unruh detector in circular motion would nevertheless show a signal. 
This puzzling conflict in the conclusions between Refs.\,\cite{Denardo:1978dj, Letaw:1979wy} for the vacua 
and Refs.\,\cite{Letaw:1979wy, Bell:1982qr, Bell:1986ir} for nonzero signals in Unruh detectors is immediately resolved by reasoning via Poincar\'e invariance.
Namely, the analyses which purport to show the existence of a RUE were conducted in frameworks that break rotational symmetry.  
In those studies, the kinematical operators of rotation become irrelevant, being supplanted by boost operators. 
Consequently, pseudodynamical phenomena, including the Unruh effect, appear.

We first review the initial studies \cite{Denardo:1978dj, Letaw:1979wy}, which concluded that there is no RUE. 
The first investigation of the RUE was performed by Denardo and Percacci \cite{Denardo:1978dj}. 
Their calculation uses IF dynamics and explicitly preserves cylindrical symmetry by adopting cylindrical coordinates for a frame whose origin is the center of the circular orbit. 
In addition, cylindrical waves are used for the mode expansion 
of the field. 
Thus, the IF rotation operators are relevant and, in accord with their kinematical nature, no RUE was found. 
Furthermore, Ref.\,\cite{Denardo:1978dj} they observed that with cylindrical waves, the Minkowski and accelerated modes are identical; and thereby 
arrives, via explicit calculations, to the same conclusion that one reaches from symmetry considerations.

The next RUE study was completed in Ref.\,\cite{Letaw:1979wy}, seemingly without knowledge of the Ref.\,\cite{Denardo:1978dj} results. 
It used the same coordinate system. 
Thus, the calculation \cite{Letaw:1979wy} of the number operator, $\hat N$, preserves the cylindrical symmetry of the system, and it was again found  that
\begin{equation}
\hat N_{\rm rot.frame} \ket{0}_{\rm Minkowski} = 0\,, 
\tag{A}
\label{Eq1}
\end{equation}
\textit{viz}.\ the vacua of the rotating and inertial frames are identical, signalling the absence of a RUE.

It was recognized why the RUE is absent in the analyses of Refs.\,\cite{Denardo:1978dj, Letaw:1979wy}.
Namely, there is no spacetime mixing: $t=t^\prime$, where $t$ is the time in the inertial frame and $t^\prime$ is the time in the rotating frame whose origin coincides with the center of rotation \cite{Fulling:2014wzx}. 
This evidently agrees with IF rotations being  kinematical operators and the central argument in~Ref.~\cite{Deur:2024szw}  
that the IF space-time mixing generates pseudodynamics, including the Unruh effect. 

However, with the cause behind the absence of a RUE in Refs.\,\cite{Denardo:1978dj, Letaw:1979wy} identified, it was nevertheless proposed, in order to create a RUE, to adopt a frame where spacetime mixes, {\it e.g.}, $t^\prime = \gamma(t+\Omega r^2 \phi), \phi^\prime =\gamma(\phi+\Omega t)$, where $\phi$ is the azimuthal angle, $r$ is the radius,
$\Omega$ is the angular velocity, and $\gamma$ is the Lorentz factor \cite{Fulling:2014wzx}. 
This choice introduces dynamics in a transformation, thereby violating Poincar\'e invariance: an IF rotation must be kinematical, as Dirac showed \cite{Dirac:1949cp}.
Thus, pseudodynamics --interpreted as a RUE-- appears by forcing dynamics into a transformation which should be kinematical. 

Using such a transformation is not strictly incorrect; it is a matter of practitioner choice, and can be justified if one attaches the frame to the orbiting detector, {\it i.e.}, the frame origin moves with respect to the rotation center and there is no cylindrical symmetry in that frame. 
However, in that case, it is incumbent upon the practitioner to include a complete analysis of all pseudoeffects and their compensating influences.

Clearly, studies of systems which do not display rotational symmetries are irrelevant to the points we are canvassing herein; namely, that there is no Unruh effect when the transformation relevant to the acceleration is a kinematical operation. 
However, it is still useful to inspect such studies in order to identify how the pseudodynamics arises and to examine the claim of its observability. 

We begin with Ref.\,\cite{Letaw:1979wy}. 
After arriving at the result in Eq.\,\eqref{Eq1}, Ref.\,\cite{Letaw:1979wy} calculated that, despite this fact, an accelerated detector would somehow show a signal, although it was not understood why at the time. 
From the perspective of Ref.\,\cite{Deur:2024szw}, the reason is clear: the calculation addressing the detector behavior implicitly breaks cylindrical symmetry
--see Ref.\,\cite[Eq.\,(37)]{Letaw:1979wy}-- thereby allowing for a pseudoeffect, which was interpreted as a detector response. 

However, the influential studies concluding the existence of a RUE 
are those described in Refs.\,\cite{Bell:1982qr, Bell:1986ir}. 
Therein, the argument for concluding a positive detector response 
is the same as that of Ref.\,\cite{Letaw:1979wy}: a local comoving frame attached to the orbiting detector (consisting of a spin-1/2 charged particle) is adopted. 
This leads to a proper time $\tau=t/\gamma - \beta y^\prime$ mixing of time and space, which thus introduces pseudodynamics. 
(This expression for $\tau$, with $y^\prime$ singled out, also shows the violation of cylindrical symmetry around the $z$-axis.) 
Boosts are also involved in Refs.\,\cite{Bell:1982qr, Bell:1986ir} as the motion of the accelerated frame (more exactly, the succession of inertial instantaneous comoving frames) is decomposed into a time translation, a rotation around the $z$-axis, and a boost. 
The term involving the rotation operator vanishes in the derivation of the detector response, 
as expected from the previous discussion (no RUE, even using IF dynamics), whereas the influence of the boost remains (linear Unruh pseudoeffect within IF dynamics). 
Thus, Refs.\,\cite{Bell:1982qr, Bell:1986ir} are consistent with the discussion in Ref.\,\cite{Deur:2024szw}; and the presence of an IF boost makes it unsurprising that an Unruh effect is identified in Ref.\,\cite{Bell:1986ir}. 

However, what seemingly contradicts Ref.\,\cite{Deur:2024szw} is that this effect is asserted in Refs.\,\cite{Bell:1982qr, Bell:1986ir} to not only be observable but to have already been observed, albeit indirectly. 
This claim has been pivotal in establishing a belief in the reality of an Unruh effect \cite{Fulling:2014wzx}. 
Therefore, in the next section, we will examine the study of Ref.\,\cite{Bell:1986ir} in detail in order to identify the origin of this claim and assess whether it truly contradicts the conclusions in Ref.\,\cite{Deur:2024szw}.

\section{Claims of the existence of a rotational Unruh effect  and its evidence \label{Bell-Leinaas}}
\subsection{Interpretation of the Sokolov-Ternov effect as arising from the Unruh effect}
It is well-established and experimentally verified that there is a theoretical limit for the natural polarization of electrons in storage rings. 
This is the Sokolov-Ternov effect \cite{Sokolov:1963zn}. 
It arises when particles, in practice electrons or positrons, emit synchrotron radiation. 
The effect has a theoretical maximum polarization of $P_{\rm max}= 8/5\sqrt 3 \simeq 0.924$.
This limit was first derived semi-classically \cite{Sokolov:1963zn} and confirmed by subsequent studies  \cite{Derbenev:1973ia, Jackson:1975qi}.
All these studies were conducted before the Unruh effect was suggested \cite{Unruh:1976db}.

References~\cite{Bell:1982qr, Bell:1986ir} 
interpreted the fact that $P_{\rm max} < 1$ as a manifestation of a RUE. 
Simply  stated, the idea is that particles with nonzero spin in a magnetic field (which in this case ensures the orbit of the particles in the storage ring) are fully polarized at a temperature $T=0\,$K because the magnetic field has lifted the energy degeneracy between the $\ket{\uparrow}$ and $\ket{\downarrow}$ spin states. 
At $T=0\,$K, all particles are, after a transition time, in their ground state, $\ket{\uparrow}$ for electrons; 
but when $T>0\,$K, thermal energy allows electrons to be excited to the state $\ket{\downarrow}$, thereby reducing the polarization. 
Then, Refs.\,\cite{Bell:1982qr, Bell:1986ir} interpret $P_{\rm max}$ as owing to a large Unruh temperature, which above $10^3\,$K for a standard storage ring. 

Notably, in Refs.\,\cite{Bell:1982qr, Bell:1986ir}, the Unruh temperature is not directly used to compute the consequent $P_{\rm max}$. 
Rather, $P_{\rm max}$ is obtained from the spin-flip transition rate that owes to quantum fluctuations of the field vacuum.%
\footnote{In a storage ring, a magnetic field $\vec B$ is required to ensure the orbit and focus of the beam. 
In the comoving frame used in Ref.\,\cite{Bell:1986ir}, the magnetic field also induces an electric field, $\vec E$. 
Thus, the term ``vacuum'' quantum fluctuations may appear misleading,
since the fluctuations are in addition to already present fields. 
Regardless, the reason why particle loops in the FF vacuum are forbidden still applies, namely that: FF momenta are always positive; momentum is conserved; and the vacuum has zero momentum, ${\mathsf P}_{\rm field} = 0$. 
The fields in Ref.\,\cite{Bell:1986ir} have stationary classical parts on the classical orbit, with thus field polarization $P_{\rm field} = 0$. 
Thus, the field fluctuations may be considered as ``vacuum'' fluctuations and associated with an Unruh effect.} 
The mean energy of the fluctuations is then computed and, assuming that they are thermally distributed, an Unruh temperature is deduced. 
It has the same form as an Unruh temperature, but for a numerical factor.

Since Refs.\,\cite{Bell:1982qr, Bell:1986ir} could interpret the Sokolov-Ternov effect in terms of an Unruh effect, the former is now considered a manifestation of the latter \cite{Fulling:2014wzx, Mane:2005xh}. 
However, as already pointed out in~\cite{Bell:1986ir}, $P_{\rm max}$ is not a direct proof of the Unruh effect because, at the same time as it affects particle polarization, it becomes irreducibly intertwined with another phenomenon, namely the feedback of the particle's spin-flip on its orbit, and one effect cannot exist independently of the other. 

In the next subsection, we follow the calculation of Ref.\,\cite{Bell:1986ir} and show how the RUE actually arises from the {\it linear} Unruh effect \cite{Fulling:1972md, Davies:1974th, Unruh:1976db}; a RUE proper does not arise, even in IF treatments, consistent with the IF rotation operators being kinematical.

\subsection{A derivation of the Sokolov-Ternov effect}

Here we list some of the key steps in the Ref.\,\cite{Bell:1986ir} derivation \cite{Bell:1986ir} of $P_{\rm max}$. 
While they may not seem critical in the context of 
\cite{Bell:1986ir}, they are material to the present analysis of the origin of the RUE. 
They are discussed in the context of IF pseudodynamics in order to identify the nature of the mechanisms considered in Ref.\,\cite{Bell:1986ir} 
and whether that calculation provides a complete IF description. 
Henceforth, for cross-referencing convenience, our numerical equation numbering corresponds to that of Ref.\,\cite{Bell:1986ir}, \textit{viz}.\ each of our equation numbers refers directly to the same numbered equation in that source. 
Equations labelled alphabetically are specific to this manuscript.
Furthermore, to aid with tracking pseudodynamics, we use colored fonts in the equations for terms whose origin is 
{\color{ForestGreen}time translations (green)},
{\color{red}boosts (red)},
{\color{RawSienna} orbital angular momentum (OAM, brown)},
{\color{BurntOrange} spin (orange)},
the part of the anomalous magnetic moment (AMM) {\color{Turquoise} which does not involve a boost (cyan)}, {\color{blue} the part which involves a boost (blue)}, 
and {\color{violet} the terms which mix {\color{red}boost}, and  {\color{blue}AMM involving a boost} (violet)}. Thus, to track the pseudodynamics, one can follow the {\color{blue}blue} 
and {\color{red}red} terms, and their {\color{violet}violet} mix. 

\begin{figure}[t]
 \center
\includegraphics[width=0.4\textwidth]{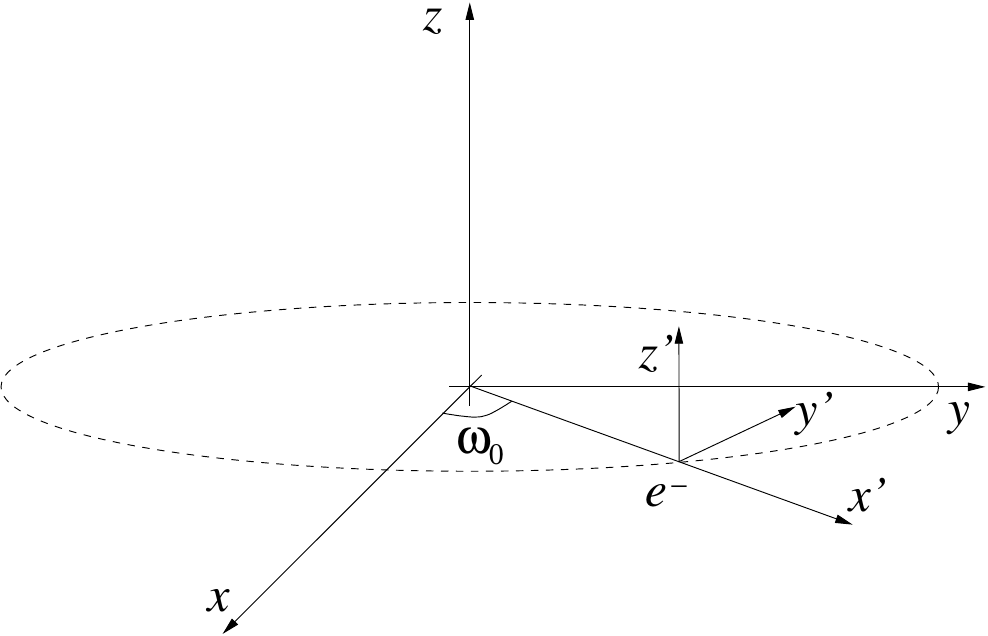}
\caption{Laboratory frame $(t,x,y,z)$ and comoving frame $(\tau,x',y',z')$, with $\tau$ the proper time. The circle depicts a classical particle orbit. 
\label{fig:frame_BL}
}
\end{figure}

The laboratory and comoving frames are depicted in Fig.\,\ref{fig:frame_BL} with primed quantities indicating a vector in the comoving frame.
The orbit of electrons in a storage ring is cylindrically symmetric around the $z$-axis. 
The $(\tau,x',y',z')$ frame explicitly violates this symmetry, as is visually evident, but is also clear from the proper time  
$\tau=t/\gamma - \beta y' $, 
where $y'$ is present but not $x'$.

The operator evolving the particle state from one inertial comoving frame at $\tau$ to the next one at $\tau+\delta \tau$ is:
\begin{eqnarray}
H^\prime = {\color{ForestGreen}H} + {\color{RawSienna} \frac{a^\prime}{\beta c} J_z} + {\color{red}\frac{a^\prime}{c}K_x}~,  \setcounter{equation}{5}
\label{eq:B-L.Eq.5}
\end{eqnarray}
where {\color{ForestGreen}$H$} is the Hamiltonian for the electron in an electromagnetic field, \textit{viz}.\ the generator for time translations,
{\color{RawSienna}$J_z$} is the generator for rotations around $z=z^\prime$,
{\color{red}$K_x$} is the boost generator in the $x$-direction,
$a^\prime$ is the acceleration in the co-moving frame, and
$\beta$ is the ratio of the velocity to the speed of light, $c$. 
Equation~(\ref{eq:B-L.Eq.5}) shows that the circular motion of the electron contains three distinct parts:
(\textit{I}) a time evolution part, given by {\color{ForestGreen}$H$};
(\textit{II}) a rotation of the co-moving frame around the $z^\prime$ axis, given by {\color{RawSienna}$J_z$}; 
and (\textit{III}) a change of linear velocity of the origin of $(\tau,x',y',z')$, given by {\color{red}$K_x$}.
The explicit expression for $H^\prime$ is:
\begin{eqnarray}
H' =  && \hspace{-0.4cm} (1-a^\prime x^\prime/c^2)\big[{\color{ForestGreen} \hat \beta mc^2 +\vec \alpha . (\vec{p'} - \frac{e}{c}\vec A^\prime)c +e\varphi^\prime } \nonumber \\
    &&+    \frac{(g-2)}{2} \frac{e \hbar}{2mc}({\color{blue} i \hat \beta \vec \alpha . \vec E^\prime} - {\color{Turquoise} \hat \beta \vec \sigma . \vec B^\prime } ) \big] \nonumber \\
    &&+   {\color{red}\frac{i \hbar a^\prime}{2c} \alpha_x} \nonumber \\
    &&-   \frac{a^\prime}{\beta c} ({\color{RawSienna}L^\prime_z} +{\color{BurntOrange}\frac{\hbar}{2} \sigma_z})~, \setcounter{equation}{6}
\label{eq:B-L.Eq.6}
\end{eqnarray}
where $\hat \beta$, $\vec \alpha$ are the Dirac matrices;
$\vec \sigma$ are the Pauli matrices;
$m$, $e$, and $g$ are the electron mass, charge, and AMM; 
$\vec {p^\prime}$ is the momentum operator; 
($\varphi^\prime, \vec A^\prime$) and ($\vec E^\prime, \vec B^\prime$) are the electromagnetic potentials and fields; 
and {\color{RawSienna}$L^\prime_z$} is the operator for the OAM along $z^\prime$.
Line 2 of Eq.\,\eqref{eq:B-L.Eq.6} is not present in Eq.\,\eqref{eq:B-L.Eq.5} and was added to account for the electron AMM coupling with $(\vec E^\prime, \vec B^\prime)$.
Line 3 shows the boost term in the $x$-direction, since  {\color{red}$\vec K =i\gamma^0\vec \gamma=i \vec \alpha$}. 
Finally, line 4 displays the rotation operator separated into OAM and spin contributions.

In Eq.\,\eqref{eq:B-L.Eq.6}, line 1 is the genuine time evolution operator {\color{ForestGreen}$H$}, which does not induce pseudodynamics, by definition.  
Likewise, {\color{RawSienna}$L^\prime_z$} and {\color{BurntOrange}$\sigma_z$} in line 4 are rotation operators, which are kinematical in the IF. 
Therefore, they are also not sources of pseudoeffects.  

On the other hand, the IF boost operator, {\color{red}$i \alpha_x$} on line 3, is dynamical and thus induces pseudoeffects.  
The AMM term in line 2 does not reflect any symmetry of spacetime, \textit{i.e}., it is not one of the 10 Poincar\'e generators. 
It contains a dynamical operator, the boost {\color{blue}$i \vec \alpha = \vec K$}, and a kinematical one, the spin operator {\color{Turquoise} $\vec \sigma$}.  
Therefore, the AMM term added to the Hamiltonian is a partly dynamical term that balances the pseudodynamics from the IF boost line 3. 
This parallels the need for a pseudodynamical spin-orbit term in the calculation of the Gerasimov-Drell-Hearn (GDH) relation \cite{Gerasimov:1965et, Drell:1966jv} for composite systems to balance the pseudoeffects of the IF boost \cite{McGee:1967zza, Brodsky:1968ea, Brodsky:2022fqy}, lest this empirically verified relation, which reflects  
basic QFT foundations \cite{Helbing:2006zp}, is violated.  
Therefore, as in the IF analysis of Ref.\,\cite{Brodsky:1968ea}, one may anticipate that the IF analysis in Ref.\,\cite{Bell:1986ir} is complete.  

Continuing with the derivation in Ref.\,\cite{Bell:1986ir}, a Foldy-Wouthysen transformation is applied to $H^\prime$, whose spin part then becomes:
\begin{eqnarray}
H_{\rm spin} =
    &-&\frac{e\hbar}{4mc}\vec \sigma \big[{\color{Turquoise}(1-a^\prime x^\prime/c^2) g\vec B^\prime} +  {\color{blue}\frac{1}{mc} (g-1) \vec E^\prime \times \vec \pi'} \big]  \nonumber \\
    && + {\color{red}\frac{1}{4}\frac{a^\prime \hbar}{m c^2}( \sigma_y \pi^\prime_z - \sigma_z \pi^\prime_y )} \nonumber \\
    && - {\color{BurntOrange}\frac{1}{2}\frac{a^\prime \hbar}{\beta c} \sigma_z}~, \setcounter{equation}{17}
\label{eq:B-L.Eq.17}
\end{eqnarray}
where $\vec \pi^\prime$ is the generalized momentum operator. 
Since only the spin part is considered, {\color{ForestGreen}$H$} 
and 
{\color{RawSienna}$L_z^\prime$},  lines 1 and 4 of Eq.\,\eqref{eq:B-L.Eq.6}, respectively, are ignored. 
The absence of {\color{RawSienna}$L_z^\prime$} is notable. 
It is the operator relevant to the RUE, just as the boost operator is the one relevant to the linear Unruh effect. 
Therefore, it is already clear that there is no genuine RUE, even in Ref.\,\cite{Bell:1986ir}.  
This outcome is in agreement with the kinematical nature of IF rotations operators; the discussions in Section~\ref{spacetime-sym-and-Unruh-effect} and Ref.\,\cite{Deur:2024szw}; and the results of Refs.\,\cite{Denardo:1978dj, Letaw:1979wy}.  

The terms present in Eq.~(\ref{eq:B-L.Eq.17}) are the AMM (line 1), the boost in the $x$-direction (line 2), and the spin operator (line 3).  
The latter, being a kinematical operator, cannot contribute to the spin-flip probability that determines $P_{\rm max}$.  
This is demonstrated below. 
Likewise, the first term in line 1, 
{\color{Turquoise}
\begin{equation}
    -\frac{e\hbar}{4mc}\vec \sigma \big[(1-a^\prime x^\prime /c^2) g\vec B^\prime \big]\,,
    \tag{B}
\end{equation}
} 
\hspace*{-0.2\parindent}which involves no boost, does not contribute, whereas a contribution does emerge from the second term 
{\color{blue}
\begin{equation}
    \big[ -\frac{e\hbar}{4mc}\vec \sigma \frac{1}{mc} (g-1) \vec E' \times \vec \pi' \big]
    \tag{C}
\end{equation}}
which arises from the boost $i\vec \alpha$.

A ``frequency vector'' $\vec \omega = \vec \omega_1 + \delta \vec \omega$ is defined from Eq.\,\eqref{eq:B-L.Eq.17} to recover the form of a Zeeman interaction term, $H_{\rm spin} := \frac{1}{2}\hbar \vec \omega \cdot \vec \sigma$. 
The $\vec \omega_1$ part is of classical origin:
\begin{eqnarray}
\vec \omega_1 = {\color{Turquoise} - \frac{g e}{2mc} \vec B^\prime_c}
{\color{BurntOrange} - \frac{a^\prime}{\beta c} \vec u_z}\,,
\setcounter{equation}{19}
\label{eq:B-L.Eq.19}
\end{eqnarray}
with the first and second terms coming from the AMM part not involving boosts and the spin term in Eq.~\eqref{eq:B-L.Eq.17}, respectively. 
$\vec B^\prime_c$ is the classical part of $\vec B^\prime$, and $\vec u_z$ is the unit vector pointing in the $z$-direction. 
The quantum part is: 
\begin{eqnarray}
\delta \vec \omega =
&-& \frac{e}{2mc}\big\{ {\color{Turquoise} g\delta \vec B^\prime - (a^\prime x^\prime/c^2)g \vec B^\prime_c}  \nonumber \\
&& - {\color{violet} (g-2)\frac{a'}{ec} \vec u_{x'} \times \vec \pi'} \big\}\,,
\setcounter{equation}{25}
\label{eq:B-L.Eq.25}
\end{eqnarray}
where $\vec u_{x^\prime}$ is the unit vector in the $x^\prime$-direction, and $\delta \vec B^\prime$ is the quantum fluctuation of $B^\prime$. 
In Eq.~\eqref{eq:B-L.Eq.25}, the first line originates from the boostless AMM part [Eq.\,\eqref{eq:B-L.Eq.17}, line 1 with cyan fonts], 
whereas the last term [second line] comes from a mix of boost [Eq.\,\eqref{eq:B-L.Eq.17}, line 2] and the AMM part involving a boost [Eq.\,\eqref{eq:B-L.Eq.17}, line 1 with blue fonts]. 

From $\vec \omega$, the transition rate for spin-flip up (+) or down ($-$) of an electron over a macroscopic time $T$ is:
\begin{eqnarray}
\Gamma_{\pm} = \frac{1}{4T} \bigg| \int_{-T/2}^{T/2} d\tau ~ \exp(\pm i \omega_1 \tau) \delta \omega_{\mp} \ket{0} \bigg|^2 ~,
\setcounter{equation}{22}
\label{eq:B-L.Eq.22}
\end{eqnarray}
where the meaning of $\ket{0}$ is not important here\footnote{It is the state of the unperturbed radiation field.} 
and
$\delta \omega_\mp := \delta \omega_x \mp i \delta \omega_y$. 
The spin term, line 3 in Eq.\,\eqref{eq:B-L.Eq.17}, contributes only to the classical part, $\vec \omega_1$, and not to $\delta \vec \omega$, which dictates the electron spin-flip probability. 
Thus, as anticipated and as in the case of the OAM, the spin term 
does not enter into an interpretation of the Sokolov-Ternov limit in terms of the Unruh effect.

Using the Heisenberg equation of motion, the $\pm$ components of Eq.\,\eqref{eq:B-L.Eq.25} yield: 
\begin{eqnarray}
\delta \omega_\mp = 
&-& \frac{e}{2mc}\big\{ {\color{Turquoise} g  B^\prime_{q\mp} }  \nonumber \\
&& + {\color{violet} 2\beta E^\prime_{qz}}  \nonumber \\
&& + \frac{1}{\beta}({\color{Turquoise} g} - {\color{violet} 2\beta^2}) \delta E^\prime_{c z}  \big\}~,
\setcounter{equation}{31}
\label{eq:B-L.Eq.31}
\end{eqnarray}
with the index $q$ indicating the quantum part of the fields. 
The term in line 1 originates from the boostless part of the AMM. The term in line 2 is from both the boost and the AMM 
part involving a boost. In line 3, the term ${\color{Turquoise} g\delta E^\prime_{c z}/\beta}$ stems from the boostless AMM part, whereas the other is 
a mix of a boost and the AMM part with a boost. 

As the $\delta \omega_\mp$ calculation proceeds, the boost and the AMM part involving a boost keep mixing and, in the last term of Eq.\,\eqref{eq:B-L.Eq.31}, the boostless term {\color{Turquoise}$\propto g$} drops out. 
Eventually, 
\begin{eqnarray}
\delta \omega_\mp = - \frac{e}{2mc}\big\{ {\color{Turquoise} g B'_{f^{\mp}}}   + {\color{violet} 2E'_{fz}}  \big\},
\setcounter{equation}{45}
\label{eq:B-L.Eq.45}
\end{eqnarray}
where $B'_{f^{\mp}}$ and $E'_{fz}$ are particular parts of the electromagnetic field. 
The first term in Eq.~(\ref{eq:B-L.Eq.45}) is from the boostless part of the AMM and is thus kinematical.
The second term comes from both the boost and the AMM 
part involving a boost (as noted, the boostless term {\color{Turquoise}$\propto g$} in line 3 of Eq.~(\ref{eq:B-L.Eq.31}) has dropped out) 
and is pseudodynamical.  As expected, neither the OAM nor the spin operators in Eq.~(\ref{eq:B-L.Eq.6}) are present in Eq.~(\ref{eq:B-L.Eq.45}).
(Note that, in Eq.\,\eqref{eq:B-L.Eq.45}, $q\to f$ because Ref.\,\cite{Bell:1986ir} neglected the dissipative part of the fields, so ``quantum'' becomes ``free'' field;
$\beta \approx 1$ was also assumed therein; 
and the final term  of Eq.\,\eqref{eq:B-L.Eq.31} was discarded as practically negligible.)

The spin-flip rates $\Gamma_\pm$ 
can now be calculated, and from them, the electron polarization in the ring, 
$P=\frac{\Gamma_+ - \Gamma_-}{\Gamma_+ + \Gamma_-}$. 
Crucially, if {\color{violet} $E^\prime_{fz}\to 0$}, \textit{i.e}., the boost-induced term containing pseudodynamics is absent, then the depolarization is not significant: $P_{\max} \simeq 0.981$ \cite{Bell:1986ir}. 
Only when {\color{violet} $E^\prime_{fz}$} is included, does
$P_{\max} \simeq 0.924$, which demonstrates that without introducing pseudodynamics via boosts, there is no rotational Unruh effect.
This fact is in accordance with the conclusion of Section~\ref{spacetime-sym-and-Unruh-effect} and the general discussion in Ref.\,\cite{Deur:2024szw}. 

The result of Ref.\,\cite{Bell:1986ir}, derived in the comoving frame, 
agrees with that of Ref.\,\cite{Jackson:1975qi}, derived in the laboratory frame with cylindrical symmetry. 
In the latter, the classical Thomas precession causes $P_{\rm max}  \simeq 0.924$. 
Thus, it is unrelated to any Unruh effect, the existence of which has always been argued to be of a purely quantum nature. 

The last steps in Ref.\,\cite{Bell:1986ir} are to compute the energy $\mathcal{E}$ contained in the field fluctuations, demonstrate that it approximately follows a thermal spectrum, and deduce a temperature $T=\mathcal{E} /k$. 
Aside from a factor of $\frac{48}{13 \pi \sqrt3} \simeq 0.68$, $T$ has the same expression as an Unruh temperature for linear accelerations.  
The authors conclude that this concordance supports interpreting $P_{\rm max}$ as owing to an Unruh effect,
while cautioning that this is not a direct proof of the existence of the Unruh effect, because it is irreducibly entangled with the spin-flip feedback on the electron beam in the ring.

\subsection{Discussion of the IF approach}

In addition to the linear Unruh effect, caused by the third term (red fonts) in Eq.\,\eqref{eq:B-L.Eq.6}, another pseudoeffect arises from the second part (blue fonts) of the line 2 term in Eq.\,\eqref{eq:B-L.Eq.6}. 
The two pseudoeffects thoroughly mingle while excluding any kinematical term. 
Their coupled effect, in {\color{violet} violet} fonts in the subsequent equations, agrees with the standard analyses \cite{Jackson:1975qi, Derbenev:1973ia} of the Sokolov-Ternov effect. 
Since these analyses account for the system's symmetry by using cylindrical coordinates centered on the ring's symmetry axis, and since they are essentially classical and hence unrelated to the Unruh effect, the agreement indicates that the Ref.\,\cite{Bell:1986ir} analysis is complete,   
with all the pseudoeffects included. 

As has been shown, all the pseudodynamics eventually stem from boosts. 
In fact, it cannot be otherwise since only the boost operators are dynamical in IF dynamics (besides the Hamiltonian). 
Therefore, there is no RUE, and the effect discussed in Ref.\,\cite{Bell:1986ir} traces to the linear Unruh effect \cite{Unruh:1976db}. 
It emerges in Ref.\,\cite{Bell:1986ir} because the analysis requires boosts to describe the system using frames that do not respect its rotational symmetry.

Although Ref.\,\cite{Bell:1986ir} provides a complete and correct analysis, this does not imply that the Unruh effect is real. 
Drawing again on the analogy with classical dynamics, a complete analysis in a non-Galilean frame 
would yield the same results as the analysis done in an inertial frame, but evidently without implying that pseudoforces are real phenomena.
Using another example, the IF analysis in Ref.\,\cite{Bell:1986ir} is comparable to the IF analysis of the GDH relation in Ref.\,\cite{Brodsky:1968ea}: 
both IF analyses include several pseudoeffects, which combine to reach a correct final answer. 
If some of the pseudoeffects are overlooked, the answer is erroneous, \textit{e.g}., Ref.\,\cite{Barton:1967at} for GDH or Refs.\,\cite{Hawking:1975vcx, Fulling:1972md, Davies:1974th, Unruh:1976db} for boost pseudodynamics.

\section{Summary and Conclusions}

Once a spacetime symmetry is violated, its associated conservation law is voided, {\it e.g.,} momentum conservation for non-Galilean frames. 
Then, pseudodynamics, like the centrifugal force in Newtonian dynamics, arises to restore the lost conservation law. 
The nonrelativistic Galilean invariance of space generalizes in the relativistic case to the Poincar\'e invariance of spacetime. 
Dirac showed that among the several possible ways to analyze a relativistic system, depending on time definition \cite{Dirac:1949cp}, the front form (FF) breaks Poincar\'e invariance the least because the FF has the smallest number of dynamical operators. 
Furthermore, the two operators that would create pseudodynamics, two rotation operators, are typically irrelevant to the description of a system. This effectively makes a FF analysis Poincar\'e-invariant, {\it i.e.}, free of pseudodynamics and ensures an objective description of the system. 

In contrast, the instant form (IF) operators that generate pseudodynamics are the boosts, which are generally crucial in any analysis. 
In Ref.\,\cite{Deur:2024szw}, it was revealed that pseudodynamics, originating from IF boosts, creates an Unruh effect. 
In FF, boosts are kinematical operators and no Unruh effect appears. 

An equivalent perspective is that the FF vacuum is trivial because FF quantization respects Poincar\'e invariance and causality, 
and therefore, the Unruh effect does not arise. 
Rather, it is a pseudoeffect, necessary in an IF description; just as in Newtonian mechanics, centrifugal forces are necessary when non-Galilean frames are used. 
It thus cannot be associated with an objective physical effect.

One must note that Poincar\'e invariance holds for both classical and quantum dynamics. 
Indeed, QFT, the framework of the Unruh effect, assumes Poincar\'e invariance. 
Thus, a hypothetical quantum violation of Poincar\'e invariance, \textit{viz},\ a quantum anomaly, is irrelevant when one compares the Unruh effect in IF and FF. 
In any case, there is presently no experimental evidence for a violation 
of Lorentz invariance or linear/angular momentum conservations, which together form Poincar\'e invariance.

For the same reasons that preclude an Unruh effect in FF, there should be no rotational Unruh effect in {\it both} the IF and FF, owing to the kinematical nature of the IF and FF rotation operators around $z$. 
Indeed, the IF analyses which respect the cylindrical symmetry find that the inertial and Rindler vacua are identical \cite{Denardo:1978dj, Letaw:1979wy}. 
However, other IF analyses have been conducted in frames not reflecting the cylindrical symmetry, which then involve boosts \cite{Bell:1982qr, Bell:1986ir} that generate a linear Unruh effect. 

If the IF frame breaks cylindrical symmetry, a complete analysis, including all pseudodynamics, should nevertheless yield no observable Unruh effect \cite{Deur:2024szw}. 
We argue that such a correct and complete IF analysis is performed in Ref.\,\cite{Bell:1986ir}, wherein another pseudoeffect, arising from a ($g-2$) term, is accounted for. 
The boost and ($g-2$) pseudoeffects thoroughly merge to become inseparable, and the final result agrees with standard semiclassical 
analyses \cite{Sokolov:1963zn, Derbenev:1973ia, Jackson:1975qi}. 
These do not invoke the Unruh effect in their interpretation but, instead, the Thomas precession.
This is a classical phenomenon unrelated to the Unruh effect, which is quantum in essence.

The analysis in Ref.\,\cite{Bell:1986ir} exemplifies how, when violating a spacetime symmetry, pseudodynamics arise to compensate for the violation.
Using an analogy with Newtonian dynamics, a complete analysis in a non-Galilean frame, involving centrifugal, Coriolis, and Euler pseudoforces, 
would be correct and would agree with an analysis done in an inertial frame. 
Yet, this does not mean that centrifugal, Coriolis, or Euler forces, or the Unruh effect, are real phenomena. 
In other words, while the treatment in Ref.\,\cite{Bell:1986ir} and its interpretation in terms of an Unruh effect are correct, the interpretation is neither fundamental nor objective. 

In the description of the static Earth-Moon system (ignoring their finite size) in a frame attached to one of the bodies, attributing their staticity to the exact balance between gravitational and centrifugal forces is not fundamental. 
The fundamental explanation relies on the gravitational force and 
linear momentum conservation: as has been said \cite{Feynman:1963uxa}: 
the Moon continuously falls on Earth but keeps missing it due to its sideways momentum. 
Choosing a non-Galilean frame voids momentum conservation, which needs to be restored by a non-fundamental effect, the centrifugal force. 

The question of the existence of an Unruh effect owing to centripetal acceleration, which should have an immediate and obvious answer, has turned out to be perplexing when addressed with the standard IF approach \cite{Fulling:2014wzx}. 
Yet, the answer is immediate from the perspective of Ref.\,\cite{Deur:2024szw}, which considers the nature of the transformations involved in the effect, namely rotations (inducing no effect) and, optionally \cite{Bell:1986ir}, boosts (inducing pseudo-effects). 

The same considerations can also answer other variations on the Unruh effect. 
For example, one may ask if the zero-modes of the Higgs field would allow a generic field $\phi$ to produce an Unruh effect. 
The argument \cite{Deur:2024szw} that kinematic operators (FF boosts, and IF or FF rotations) cannot produce genuine dynamics shows that even if a Higgs field is added to the Lagrangian, it would not allow for the production of $\phi$ particles via an Unruh effect. 

Indeed, when investigating the Unruh effect \cite{Deur:2024szw}, the only role of the Lagrangian is to determine the dispersion relation of $\phi$. The remaining discussion is only kinematic considerations. 
The form of the dispersion relation for any field in the Standard Model, $p^2=m^2$ or $E=|p|$, is unaffected by the Higgs field. 
Therefore, adding a Higgs term to the Lagrangian does not affect the Unruh effect demonstration.\footnote{
This reasoning also immediately provides a negative answer to the related question of whether the Higgs field can produce Higgs bosons via the Unruh effect, since the Higgs field has the same standard dispersion relation, $p^2=m^2$.} 
Furthermore, it would not be kinematically permitted to have a nonzero Unruh effect arising from interactions with the zero-mode Higgs background, since the Higgs state in the FF has $p^+=0$.

\begin{acknowledgements}
This material is based upon work supported by 
U.S.\ Department of Energy, Office of Science, Office of Nuclear Physics, contract DE-AC05-06OR23177 (AD);
U.S.\ Department of Energy contract DE-AC02-76SF00515 (SJB); 
National Natural Science Foundation of China grant no.\ 12135007 (CDR);
and U.~S.~National Science Foundation award no.\ 1847771 (BT).
\end{acknowledgements}

\medskip

\noindent {\bf Financial interests}: The authors have no competing interests to declare that are relevant to the content of this article.

\bibliographystyle{elsarticle-num-names}
\bibliography{BiBUnruhR.bib}

\end{document}